\documentclass{aa}
\usepackage{graphics}
\begin{document}

   \thesaurus{05     
              (08.05.2
		11.09.1 Magellanic Clouds;
		11.19.4 NGC 330)}
   \title{Spectroscopy of Be stars in the\\Small Magellanic Cloud cluster NGC 330.}

   \author{Stefan C. Keller
          \and
          Michael S. Bessell
          }

   \offprints{S. C. Keller}

   \institute{Mount Stromlo \& Siding Spring Observatories, Private Bag, 
Weston Creek PO, ACT 2611, Australia\\
              email: stefan@mso.anu.edu.au \& bessell@mso.anu.edu.au            }
   \authorrunning{S. C. Keller et al.}
   \titlerunning{Be Stars within NGC 330}
   \date{Received $<date>$; accepted $<date>$}

   \maketitle

\begin{abstract}
The presence of an anomalously large population of Be stars in the young
cluster NGC 330 in the SMC has been noted by Feast (\cite{feast}), Bessell \&
Wood (\cite{beswood}), Grebel (\cite{greb}) and most recently Keller et
al. (\cite{kwb}). We present results from follow-up medium resolution spectra
of the bright Be stars identified in Keller et al. (\cite{kwb}) and in the
spectroscopic study of Mazzali et al. (\cite{m96}). We find that the study of
Mazzali et al. has overestimated the number of Be stars within NGC 330. Many
of the bright B type stars identified by Mazzali et al. as Be stars show no
signs of emission after correction for the diffuse HII emission pervading the
cluster field.

In the study of Mazzali et al. (\cite{m96}) evidence is presented suggesting
that all Be stars in NGC 330 have a common inclination of their rotation axes
to the line of sight. For our sample of Be stars we present the emission
equivalent widths and observed rotational velocities. An examination of these quantities shows that there is no evidence of a preferential alignment of the rotational axes, rather we are observing a population with random rotational axis alignment and disk size.


\end{abstract}

\section{Introduction}
A Be star is defined as a non-supergiant B type star (ie. luminosity class V
to III) that shows, or has at some stage shown, emission in the hydrogen
Balmer series (Jaschek et al. \cite{jaschek}). In the galactic field
population, Be stars comprise 17\% of B0-7 stars (Zorec and Briot
\cite{zb}). This proportion shows a slight maximum at spectral class B2. Stars
of this spectral type and luminosity class are necessarily young and only in
the youngest galactic clusters can we find a population residing on or near
the main sequence. The well studied young galactic clusters NGC 663, NGC 3766
and h+$\chi$ Per have ages of a few \(10^{7}\) years and have Be star
proportions of around 35-50\% of the B star population within a limited range
of luminosities of the main sequence turn-off.

The young populous cluster NGC 330 in the SMC, which is of similar age but is
more massive, has long been noted as containing a high proportion of Be
stars. A Be fraction as high as 80\% over a narrow magnitude range has been
found in our recent survey for Be stars within this cluster (Keller et al. \cite{kwb} - Paper 1). The study of this large population of Be stars, importantly of similar age, chemical composition and formation conditions, enables further insight into the Be star phenomenon.

The dominant paradigm of Be star emission is that described by the Struve
model (Struve 1931) in which line emission arises from an optically thick
rotating disk of material in the equatorial plane of the star. Much evidence
in support of this model has been found from quite diverse techniques. Wood et
al.  (1996) find the observed linear polarisation spectra for Be stars, which
arises from electron scattering in the circumstellar envelope, implies a flat
disk of material with an opening angle of 10 degrees or less. Confirmation of
a disk morphology is presented by Quirrenbach et al. (\cite{quirrenbach}) who have resolved
Be star disks with combined interferometry and polarimetry. Whilst the presence of a rotating circumstellar disk is accepted, the mechanism of its formation and maintainence remains contentious. 

In paper 1 we present the results of our survey of the cluster for
Be stars using a photometric technique which utilises the [R-H$\alpha$] and
$V$$-$$I$ colours to identify the Be stars within the cluster and the
surrounding field. We found that the fraction of Be stars to the total number
of B type stars within NGC 330 is significantly greater than that seen in
the surrounding field. In addition this population is seen to be strongly
peaked at magnitudes corresponding to the main sequence termination, whereas
the field population exhibits a constant proportion with magnitude. In Paper 1
we have proposed that the observed enhancement of Be star numbers within the
cluster can be understood as a combination of an evolutionary enhancement (due
to an increase in angular velocity during the course of main sequence
evolution) and a distribution of intrinsically high rotational velocities within the cluster. 

Mazzali et al. (\cite{m96}) present spectroscopic results from a sample of
bright B stars in NGC 330. From their sample of 14 stars, they find that 13
are Be stars, although the majority of these stars appear as normal B stars in
the photometric Be surveys of Grebel et al. (\cite{greb}) and Paper 1. It is
the primary aim of this paper is to investigate the discrepancy in Be star
identifications. To this end we have observed a sample of the brightest Be
stars identified by us in Paper 1 and those flagged as Be stars by Mazzali et al.

From the observed correlation between the equivalent width of the H$\alpha$
emission and the projected rotational velocity, $v\,sin\:i$, within their
sample, Mazzali et al. suggest that the rotational axes of the stars in NGC
330 share a common alignment and that the size of the circumstellar disc
surrounding the Be star is dependent on the stellar rotation velocity. In the
present paper we examine this assertion in the light of our revised and extended sample.

\section{Observations}

Medium resolution (0.6\AA/px) spectra of a sample of bright B and Be stars in
NGC 330 were obtained using the Double Beam Spectrograph (DBS) on the MSSSO
2.3m telescope on 23-26 September 1997. The spectra consist of two
simultaneouly recorded non-overlapping segments; blue ( 3700 - 4380\AA ) and red
( 6200 - 6800\AA ). Figure 1 shows a series of blue segments for several
emission line stars. The long DBS slit (6.3$\arcmin$) was positioned to enable spectra of several program stars to be observed at the one time. 

\begin{figure}
\resizebox{\hsize}{!}{\includegraphics{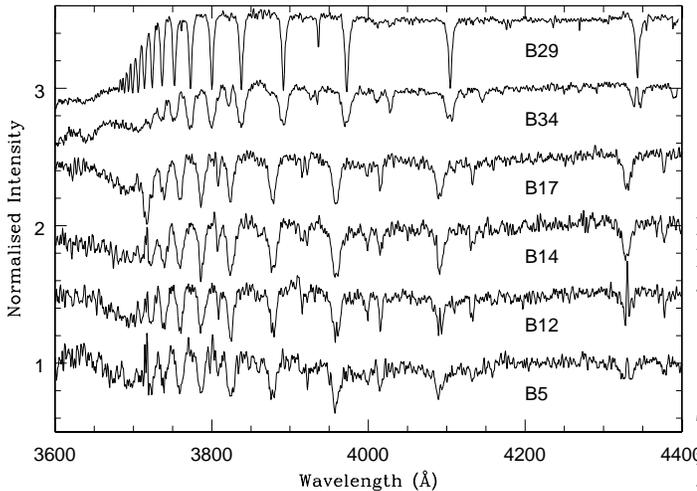}}
\caption{A sample of the spectra of Be stars considered in the present
study. B29, although an A supergiant is included here for it shows significant
H$\alpha$ emission.}
\label{specsfig}
\end{figure}

\section{Spectral analysis}
\subsection{Absorption line rotational velocities}

The 1D spectra of the individual stars were extracted from the 2D long slit
spectrograms and corrected by subtracting nearby sky. Rotational velocities
were determined by matching observed absorption lines with numerically
broadened model lines. The hydrogen Balmer lines, H$\gamma$, H$\delta$,
H$\epsilon$, were used as the basis for the profile fits. This was done so
that we could obtain rotational velocities for the fainter Be stars in our
sample in whose spectra the standard narrower lines such as HeI (4471\AA)
could not be used due to low signal to noise. A model profile was selected for
each star from Kurucz (\cite{kurucz}) grid of model atmospheres of
non-rotating stars of [Fe/H]=-0.5 dex. Surface parameters; log g and
\(T_{eff}\) were taken, where available from Mazzali et al. (\cite{m96}),
Lennon et al. (\cite{lennon}) or Caloi et al. (\cite{caloi93}). In cases when
these parameters were not available it was necessary to determine these
parameters in a coarse manner from their proximity to stars on the CMD with known parameters.

\begin{figure}
\resizebox{\hsize}{!}{\includegraphics{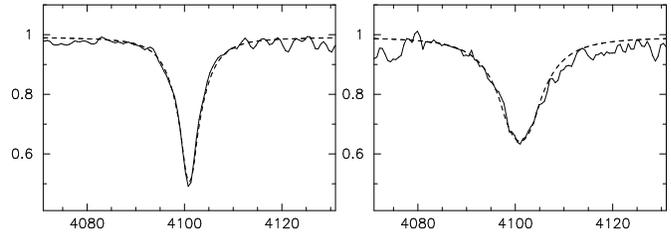}}
\caption{Examples of model fits to the H$\delta$ Balmer line of B29 (left)
with  $v\,sin\:i$=100$\rm{km\,s}^{-1}$ and B17 (right) with  $v\,sin\:i$=375$\rm{km\,s}^{-1}$.}
\label{models}
\end{figure}

To avoid the influence of emission in the Balmer series it was necessary to
fit to profiles blueward of H$\beta$. The selected model profile was then
convolved with the instrumental response function, rotationally broadened to a
given $v\,sin\:i$ (see eg. Gray \cite{gray}) and then overlayed with the
observed profile. The $v\,sin\:i$ giving the best fit to the selected line was
then found. This was repeated for H$\gamma$, H$\delta$ and H$\epsilon$ and the average $v\,sin\:i$ for each star is presented in Table 1.

\begin{figure}
\resizebox{\hsize}{!}{\includegraphics{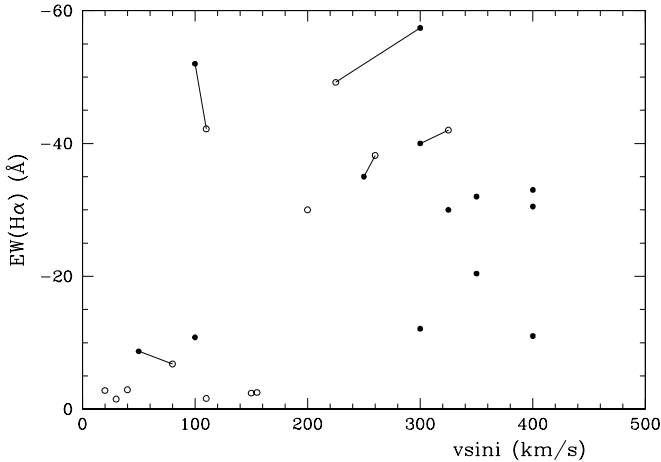}}
\caption{Plot of $v\,sin\:i$ from absorption vs. equivalent width of the H$\alpha$ emission line. The open circles are those of Mazzali et al. (1996). Those stars in common with Mazzali et al. are joined by solid lines.}
\label{figure1}
\end{figure}

The most intense emitters also show observable emission in H$\gamma$ and
H$\delta$. In these cases the fit was made to the profile ignoring the central
regions affected by emission. Out of the 32 Be stars observed in NGC 330 it
was possible to derive $v\,sin\:i$ for 13, the limitation quite simply being
the signal to noise. Indeed some of the most intense Be stars we observed were
amongst the faintest program stars and thus had very little continuum in our
exposures. The $v\,sin\:i$ results are inherently rather uncertain and we
estimate an uncertainty of $\pm$50 $\rm{km \,s}^{-1}\:$ in our results. In the
case of those stars showing strong emission this uncertainty may be slightly
higher. However if we consider those stars in common between out study and
that of Mazzali et al. we see that both arrive at similar $v\,sin\:i$ values
(see figure \ref{figure1}).

\begin{table}
\caption{Projected rotational velocities and equivalent widths for the Be
stars of the current study. Stars with a B prefix are the designations due to
Robertson (\cite{rob}). Also referenced are those according to Keller et
al. (1998). Equivalent width measures are uncertain by $\pm$2 \AA$\:$ and
\(vsini\) by $\pm$50$\rm{km\,s}^{-1}$.}
\label{tbl-1}
\begin{center}
\begin{tabular}{cccc} \hline
KWB98 & Rob.  & \(vsini\)   & W($\alpha$) \\
 &    & $\rm{km\,s}^{-1}$   & \AA \\ \hline
 
     &  B05   &  300  &  $-$40\\
     &  B12  &  100  &  $-$52\\
     &  B13  &  50   &  $-$9\\
     &  B14  &  400  &  $-$11\\
     &  B17  &  300  &  $-$12\\
     &  B21  &  250  &  $-$37\\
     &  B29  &  100  &  $-$11\\
     &  B31  &  350  &  $-$20\\
     &  B34  &  350  &  $-$32\\
     &  B35  &  300  &  $-$57\\
     &  B36  &  400  &  $-$33\\
215  & ... &  325  &  $-$27\\
235  & ... &  400  &  $-$31\\
\end{tabular}
\end{center}

\end{table}

\vspace{5mm}

\subsection{H$\alpha$ emission equivalent width}


Within the cluster environment of NGC 330 there are two effects that act to perturb our measures of equivalent width. Firstly, within the cluster field, crowding from surrounding stars acts to dilute the H$\alpha$ equivalent width. The light from extraneous background stars raises \(F_{c}\) but not (generally) the H$\alpha$ flux. This is an obvious effect in the case of all the Be stars observed towards the cluster centre. The inherent limitation is one of spatial resolution, which in the case of the present observations is 3$\arcsec$, mostly due to seeing. With this in mind, we restrict the data presented here to distances $>25\arcsec$ from the cluster core.

Secondly, NGC 330 lies in a field of diffuse, non uniform HII emission. There
are several filaments near the cluster, the brightest of which is the
elliptical region centred about B18 to the NW of the cluster centre identified
by Grebel et al. (1996) as an extension of DEM 87 (Davies et al. 1976). The nonuniformity could be readily seen in the 2D long slit spectra that sampled much of the area in and around the cluster. Particular care was taken to subtract the sky background immediately adjacent to the program stars.

\begin{figure}
\resizebox{\hsize}{!}{\includegraphics{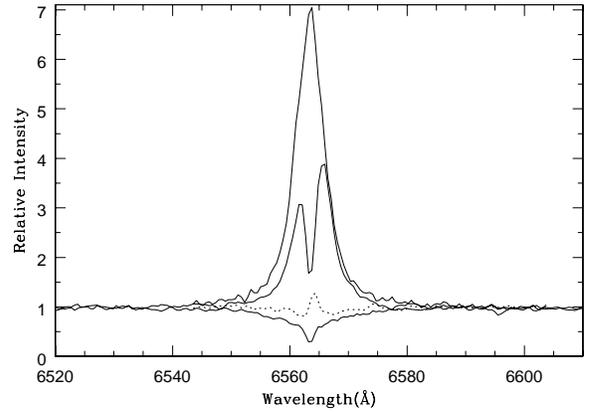}}
\caption{The H$\alpha$ profiles of stars B21 (strong emission), B31 (double peaked profile) and B16 (no emission). Indicated by the dotted line is the profile of B22 uncorrected for the surrounding HII. Note the narrow and weak emission characteristic of a HII region.}
\label{figure3}
\end{figure}

\begin{table*}
\caption{A summary of the Be status for objects near the main sequence in NGC
330. Studies refered to are as follows: G96 - the spectroscopy of Grebel et
al. (1996), obtained 1993/4; M96 - the spectroscopy of Mazzali et al. (1996),
obtained 1992; and G92 - the photometry of Grebel et al. (1992), obtained 1989. The Be status of those stars listed in Mazzali (\cite{m96}). Rob refers to the designations of Robertson (\cite{rob}).}
\label{tbl-2}
\begin{center}
\begin{tabular}{cccccclcccccc} \hline
 & \multicolumn{5}{c}{Be Status} & &\multicolumn{5}{c}{Be Status} \\
Rob. & G96 & M96 & Present & G92 & Paper 1 &Rob. & G96 & M96 & Present & G92 & Paper 1 \\ \hline
 A01 & n&y&n&n&n& A02 & n&-&n&-&n \\
 A04 & y&-&-&n&n& B04 & -&y&n&n&n \\
 B05 & -&y&y&y&y& B06 & -&y&y&y&y \\
 B07 & -&y&n&-&n& B11 & -&n&-&n&n \\
 B12 & -&y&y&y&y& B13 & -&y&y&n&n  \\
 B14 & -&-&y&y&y& B16 & n&n&n&n&n  \\
 B17 & -&-&-&y&y& B18 & -&y&-&n&n  \\
 B21 & -&y&y&y&y& B22 & n&y&n&n&n  \\
 B24 & y&y&-&n&n& B28 & -&y&n&n&n  \\
 B29 & -&n&y&-&n& B30 & n&n&n&n&n  \\
 B31 & -&-&y&-&y& B32 & -&-&n&n&n \\
 B33 & y&-&n&n&n& B35 & -&y&y&y&y  \\
 B36 & -&-&y&y&y& B37 & n&-&n&n&n \\
 B38 & n&-&n&n&n& B41 & -&-&y&y&y \\

\end{tabular}
\end{center}
\end{table*}

\section{Results}

In Table 2 we present the Be status for a sample of bright cluster members
including those in Mazzali et al. (1996). We find fewer of these stars are Be
stars than the previous spectroscopy of Mazzali et al. has indicated. As we
have mentioned previously our photometry and that of Grebel et al. (1992)
indicates that these are normal B type stars.

Within our sample two stars, B13 and B29, show emission spectroscopically but are not
flagged as Be within Paper 1. The ability of our photometric survey technique
to discern Be stars relies upon the [R-H$\alpha$] colour of the emission line
objects being significantly different to those of the majority of non-emission
objects. Photometric errors give the non-emission stars a certain dispersion
in the [R-H$\alpha$] index. This presents a limit to the equivalent width
discernable within our survey. Both B29 and B13 indeed show a comparatively small
emission equivalent width (see Table 1). Three stars from the spectroscopy of
Grebel et al. (1996) show emission undetectable to the photometric survey
technique (A4, B24 and B33). The emission equivalent widths of these stars are
unpublished. We conclude that there does not appear to be a large population
of Be stars with emission levels below that detectable in Paper 1. This is heartening for the completeness of our photometric survey technique.

Between the two observational epochs at which equivalent widths are available,
changes of up to 17\AA$\;$ in EW are evident (see figure \ref{figure1}). It is
undoubtably the case that the strength of the Be phenomenon is variable,
however it seems unlikely that we have observed the majority of these stars at an ``inopportune'' moment of no emission.

Mazzali et al. (1996) rule out the possibility of diffuse HII emission in the
cluster environment. However, as discussed above, H$\alpha$ images show that
NGC 330 indeed lies in a region of HII. We suggest that this background HII
accounts for the small intensity, narrow emission seen by Mazzali et
al. (1996). Upon investigation indeed half of the forementioned narrow
emission line stars in Mazzali et al. lie within regions of pronounced
HII. After correction for the effects of the surrounding HII emission we claim
that only 6 stars of their sample are in fact Be stars. Table \ref{tbl-2}
summarises our current knowledge of the emission status of bright MS members
of NGC 330. B29, identified by Feast and Black (1980) and Carney et al. (1985)
as an A supergiant without emission, shows significant emission in our
study. It should be noted that the previous studies have focused on the blue
portion of the spectrum where, as seen in figure \ref{specsfig}, no emission
is evident. The forementioned studies judge B29 as a field A supergiant on the
basis of its radial velocity.

Let us now discuss our results in the light of the assertion of Mazzali et
al. (1996), namely that we see, in the case of NGC 330 a common alignment of
Be star rotation axes. The argument of Mazzali et al. is laid out in
detail, however it may be summarised as follows: (a) there are no rapidly rotating stars with weak emission, nor (b) slow rotating stars with strong emission in their sample. They argue, in the case of (a) that the stars in their sample could be stars observed edge on but this would give double peaked profiles which are not observed in their sample. In the case of (b) all slowly rotating stars are observed to be weak emitters and this could be taken that we are observing stars with small disks pole on. But since the probability of seeing all such stars pole on is small, Mazzali et al. (1996) claim we must see all the stars at roughly the same $sin\:i$ between the limits of (a) and (b) (approx $\frac{\pi}{4}$). It follows from this that there must be a correlation between $v$ and W($\alpha$) (the latter being essentially a measure of the disk size)

The perceived correlation appears to arise from an unfortuitous selection of
sample stars. Our observations show that double peak profiles do occur in NGC 330, for
example, that of B31 shown in figure \ref{figure3}. Typical red to violet peak
separations for double peaked profiles in the present sample are $\sim$
2\AA$\:$ or less. The majority of Mazzali et al. (1996)'s observations were
made with a resolution of 3.6\AA $\:$and so such double peaked profiles would
have been undetectable.  Mazzali et al. made additional higher
resolution (0.6\AA) observations of 2 strong Be stars and no double peaks were
found. It turns out that the choice of two strong emitters for closer
examination was unfortuitous. As Dachs et al. (\cite{dachs}) (their figure 7) have shown, such emitters are unlikely to show double peak profiles.

Our results (solid points on figure \ref{figure1}) span a large range of
W($\alpha$) and $v\,sin\:i$ and in contrast to Mazzali et al. (1996) (their
figure 7) show no correlation between $v, sin\:i$ and equivalent width is present. This indicates that we are observing a sample of stars with $v, sin\:i$ and disc size as randomly distributed properties. There is no sign of a preferential alignment of rotational axes nor a dependence of disk size on $v$.

\section{Conclusion}

We have presented the results of our spectroscopic observations of a sample of
bright cluster members and in particular those of a number of bright Be
stars. We find that the study of Mazzali et al. (1996) has overestimated the
number of Be stars within the cluster. Of their sample of 13 Be stars we have
observed 11 of which we find
6 are in fact Be stars, the remainder are ordinary B stars. The most
likely cause of this misidentification appears to lie with the subtraction of
the strong and spatially variable HII emission within and around the
cluster. We do not find evidence for the presence of a large population
of Be stars with H$\alpha$ emission equivalent width beneath the detection threshold
of photometric surveys such as Grebel et al. (\cite{greb}) and Keller et
al. (\cite{kwb}).  

The equivalent width of emission and $v\,sin\:i$ from absorption presented by the Be stars of our sample are uncorrelated. This indicates that the stars in the Be population possess $v\;sin\:i$ and disk size of random distribution, in contrast to the assertion of Mazzali et al. (1996) that there is a general alignment of Be star rotation axes within the cluster. 

\begin{acknowledgements}

SCK acknowledges the support of an APA scholarship. We thank Eva
Grebel and Paulo Mazzali for useful comments. In the course of this research use has been made of the CDS catalogue service, the NASA Astronomical Data Service, operated by CfA, Harvard, USA and the NASA Astrophysics Data Centre.
\end{acknowledgements}

\end{document}